\documentclass[aps,prd,reprint,showpacs,nofootinbib,longbibliography]{revtex4-2}
\usepackage{amsmath,amssymb,graphicx,bm}
\usepackage{xcolor}
\newcommand{\rev}[1]{#1}
\newenvironment{revblock}{\begingroup}{\endgroup}
\usepackage[colorlinks=true,linkcolor=blue,citecolor=blue,urlcolor=blue]{hyperref}

\begin{document}

\title{Projections and topological classification of black hole thermodynamics}

\author{Di Wu$^{1,2}$}
\email{Contact author: wdcwnu@163.com, d277wu@uwaterloo.ca}

\affiliation{$^1$School of Physics and Astronomy, China West Normal University,
Nanchong, Sichuan 637002, People's Republic of China \\
$^2$Department of Physics and Astronomy, University of Waterloo, Waterloo,
Ontario N2L 3G1, Canada}

\date{\today}

\begin{abstract}
Black hole thermodynamic topology has evolved from classifications based on global topological numbers to more refined schemes including boundary behavior. This development motivates a deeper question: how are different topological manifestations connected within the complete thermodynamic structure? In this paper, we propose the thermodynamic projection conjecture, which seeks to address this question by examining different topological manifestations in the context of the complete thermodynamic structure and exploring their underlying physical connections and formation mechanisms. As a representative application of this idea, we investigate temperature-dependent thermodynamic topological phase transitions, in which the topological number varies with temperature while the physical parameters are held fixed. We examine how the complete thermodynamic structure gives rise to different topological outcomes on different temperature slices. Our results demonstrate that temperature-dependent thermodynamic topological phase transitions arise from boundary effects, whereas the pairwise generation or annihilation of equilibrium states in the interior leaves the total topological number unchanged. These results establish that, at fixed physical parameters, the topological differences observed on different temperature slices originate from the same complete thermodynamic structure and can be understood within a unified framework of thermodynamic projections. Furthermore, this study offers a new approach to understanding black hole thermodynamic structures and their rich physical behavior through the interplay between global structures and their local manifestations.
\end{abstract}

\maketitle

\section{Introduction}

Black hole thermodynamics provides a fundamental framework for understanding the physical properties of black holes and their connections to gravitational physics \cite{PRD7-2333,CMP31-161,CMP43-199,PRD15-2752}.
In particular, the study of thermodynamic stability, phase transitions, and phase structures has played a central role in revealing the rich thermodynamic behavior of black holes \cite{CMP87-577,PRD60-064018,PRD60-104026,PRD88-101502,JHEP0712033,CQG26-195011,PRD84-024037}.
Various theoretical approaches have been developed to investigate these phenomena from different perspectives, providing complementary insights into the thermodynamic properties and phase structures of black holes \cite{CQG34-063001,PRL118-021301,PRL127-091301,PRD105-106014,JHEP0822174,PRL130-181401,
JHEP1112110,JHEP0914080,JHEP1222117,JHEP0823028,
PRD103-044014,PRD102-044007,JHEP1121031,PRD100-101501,PRD105-124013,PRD108-064034,PRD108-064035,
PLB846-138227,2606.03958,2607.23644,2609.33032,2610.21830,2611.05624}.
For instance, conventional thermodynamic approaches investigate phase transitions and stability through thermodynamic potentials and response functions \cite{PRD33-2092,CMP87-577}, whereas the recently developed thermodynamic topology approach characterizes the structure of black hole equilibrium states in terms of topological invariants \cite{PRL129-191101}.
By treating black hole solutions as topological defects in thermodynamic parameter space, this approach assigns local winding numbers to individual equilibrium states and characterizes the overall thermodynamic topology through their sum, known as the global topological number \cite{PRD105-104003,PRL129-191101}.
Initially, black hole solutions were classified into three topological classes according to their global topological numbers \cite{PRL129-191101}, while a subsequent refinement extended this classification to four classes by including their thermodynamic asymptotic behavior \cite{PRD110-L081501}.
Subsequent investigations have further enriched this classification framework by identifying additional topological classes and subclasses with more diverse thermodynamic behaviors in the high- and low-temperature limits \cite{PRD111-L061501,PRD112-124024,EPJC86-187,EPJC85-1386}.
More recently, a topological subclass predicted in our previous work \cite{PRD111-L061501} has been identified in an independent study \cite{2609.20898,2610.36796}.
In parallel, extensive investigations of thermodynamic topology across a wide range of black hole systems have revealed a rich variety of topological behaviors \cite{PRD107-046013,PRD105-104053,PLB835-137591,PRD106-064059,PRD107-044026,PRD107-064023,JHEP0123102,
PRD107-106009,JHEP0623115,PRD107-024024,PRD107-084002,EPJC83-365,EPJC83-589,PRD108-084041,PLB856-138919,
PDU46-101617,PLB868-139742,PLB858-139052,CQG42-125007,PLB860-139163,PLB865-139482,EPJC85-828,2602.05231,
2605.00037,JHEP0624213}.
Together, these developments have substantially expanded our knowledge of black hole thermodynamic topology and its diverse classification patterns.

However, the growing diversity of topological classifications does not necessarily translate into a corresponding increase in our understanding of the underlying physical mechanisms.
In this respect, the increasingly detailed classification of thermodynamic topological behaviors is analogous to the analysis of individual components in classical mechanics, which provides valuable information but does not, by itself, reveal how these components are connected within the system as a whole.
Correspondingly, an analysis of the system as a whole offers a complementary perspective, motivating us to understand these diverse topological behaviors and their underlying physical connections in terms of the complete thermodynamic structure.
The central question is whether these diverse topological behaviors can be understood as different manifestations of the same complete thermodynamic structure.

Temperature-dependent thermodynamic topological phase transitions provide an important clue to answering this question, since the same black hole can exhibit different global topological numbers on different temperature slices \cite{JHEP0624213}.
This observation suggests that the global topological number may be better understood as a manifestation of the complete thermodynamic structure on a particular temperature slice, rather than simply as an intrinsic property of the black hole itself.
In this paper, we propose the thermodynamic projection conjecture, which interprets the thermodynamic topology observed on different temperature slices as different projections of the same complete thermodynamic structure.
This conjecture can be illustrated by an analogy with the passage of light through a lens or a prism.
In this analogy, the complete thermodynamic structure of a black hole corresponds to the light source, while selecting different temperature slices is analogous to viewing the light through different optical projections.
For black holes without temperature-dependent thermodynamic topological phase transitions, the temperature slices act like lenses: although different projections are selected, the resulting global topological number remains unchanged.
For black holes exhibiting such transitions, the temperature slices act like prisms: different projections reveal different global topological numbers, just as a prism reveals different colors of light, even though all these colors originate from the same source.

The key is therefore to understand how the complete thermodynamic structure determines the topological outcomes observed on different temperature slices.
To investigate this question, we focus on temperature-dependent thermodynamic topological phase transitions as a representative example of the thermodynamic projection conjecture.
We examine the equilibrium states and their topological properties on different temperature slices, with particular attention to their behavior at the boundaries of the thermodynamic domain.
Our results show that changes in the global topological number across temperature slices arise from boundary effects, while the pairwise generation or annihilation of equilibrium states leaves the total topological number unchanged.

This paper is organized as follows.
In Sec.~\ref{II}, we briefly revisit temperature-dependent thermodynamic topological phase transitions at fixed physical charges.
In Sec.~\ref{III}, we investigate the physical mechanisms underlying these transitions in three static charged AdS black holes: the four-dimensional Einstein-Maxwell-dilaton-axion (EMDA) and Horowitz-Sen (HS) black holes, and the five-dimensional Kaluza-Klein (KK) black hole.
In Sec.~\ref{IV}, we discuss our findings from the perspective of the thermodynamic projection conjecture.
Finally, Sec.~\ref{V} presents our conclusions and outlook.

\section{Revisiting Temperature-Dependent Thermodynamic Topological Phase Transitions}\label{II}

\subsection{Thermodynamic topological construction}

We first recall the thermodynamic topological construction used in the analysis below.
For a black hole with mass $M$ and entropy $S$, one introduces the generalized off-shell Helmholtz free energy
\begin{equation}
  \mathcal{F}=M-\frac{S}{\tau},
\end{equation}
where $\tau$ is an auxiliary inverse temperature.
\rev{The bath temperature is $T_b=1/\tau$; equilibrium requires $T_b=T_H$, where $T_H$ is the Hawking temperature.}
The associated vector field is defined as
\begin{equation}
  \boldsymbol{\phi}=\left(\frac{\partial\mathcal{F}}{\partial r_h},
  -\cot\Theta\,\csc\Theta\right).
\end{equation}
\rev{Here $r_h$ is the horizon radius, $\Theta\in(0,\pi)$ is an auxiliary angle, and $\phi^{r_h}$ is the radial component.}
Its zero points lie at $\Theta=\pi/2$ and satisfy $\phi^{r_h}=0$.
Solving this condition for $\tau$ gives the zero-point curve
\begin{equation}
  \tau=\tau(r_h).
\end{equation}
Each isolated zero point, indexed by $i$, carries a winding number $w_i$, and the total topological number in the relevant region is
\begin{equation}
  W=\sum_i w_i.
\end{equation}

\subsection{Temperature-dependent thermodynamic topological phase transitions at fixed physical charges}

We now revisit the temperature-dependent thermodynamic topological phase transitions of three static charged AdS black holes \cite{JHEP0624213}, keeping their physical charges and the \rev{thermodynamic} pressure \rev{$P$} fixed.
\rev{The conserved physical electric charges are obtained from the gauge-field fluxes at infinity, including scalar-dependent couplings. For the four- and five-dimensional solutions, their expressions in the normalization of Ref.~\cite{PRD84-024037} are}
\begin{revblock}
\begin{equation}
\begin{gathered}
 K_i=q_i(q_i+2m)>0,\\
 Q_i^{(4)}=\frac{\sqrt{K_i}}{2},\qquad
 Q_i^{(5)}=\frac{\pi\sqrt{K_i}}{4}.
\end{gathered}
 \label{eq:fixedQ-identification}
\end{equation}
\end{revblock}
\rev{Here $m$ and $q_i$ are the mass and charge parameters, $K_i$ is the charge combination, and $i$ labels the gauge fields. The parenthesized superscripts $(4)$ and $(5)$ on $Q_i$ denote the spacetime dimensions $D=4$ and $D=5$, respectively. Unlike $q_i$, the quantities $Q_i^{(D)}$ are conserved physical charges (see Eqs.~(2.6) and (2.15) of Ref.~\cite{PRD84-024037}).}

\rev{Since the charge normalizations are fixed, Eq.~(\ref{eq:fixedQ-identification}) gives $dQ_i^{(D)}/Q_i^{(D)}=dK_i/(2K_i)$ for $D=4,5$ and $K_i>0$. Hence $dK_i=0$ is precisely the fixed-physical-charge condition $dQ_i^{(D)}=0$. Along a family parametrized by the horizon radius, $m$ and $q_i$ must therefore vary together so as to preserve $K_i$; keeping $q_i$ alone constant would generally change the physical charge. Their dependence on the horizon radius is included when determining the zero-point curves.} Along a fixed-charge family, the first law implies
\begin{equation}
 \left.\frac{d\mathcal F}{dr_h}\right|_{Q_i,P}
 =\left.\frac{dS}{dr_h}\right|_{Q_i,P}
 \left(T_H-\frac1\tau\right).
 \label{eq:fixedQfree}
\end{equation}
\rev{This derivative relation can be checked directly for all three solutions by differentiating their masses and entropies while imposing their respective fixed-$K_i$ constraints:}
\begin{revblock}
\begin{equation}
 \left.\frac{dM}{dr_h}\right|_{K_i,P}
 -T_H\left.\frac{dS}{dr_h}\right|_{K_i,P}=0.
 \label{eq:fixedQ-firstlaw-check}
\end{equation}
\end{revblock}
\rev{The relevant fixed-charge derivatives, including the positivity of the entropy derivatives, are given explicitly with each black hole below.}
For the physical branches considered below, $dS/dr_h>0$. Hence, the zeros satisfy $\tau=1/T_H(r_h)$, and each simple root has winding number $w_i=\mathrm{sgn}(dT_H/dr_h)$. We consider positive pressure and the positive-charge branches with $r_h>0$.

\subsubsection{Four-dimensional EMDA black hole}

The four-dimensional EMDA and HS examples are obtained from the static four-charge AdS black hole in gauged supergravity \cite{NPB554-237}. The parent solution is
\begin{equation}
\begin{aligned}
 ds_4^2={}&-\mathcal H_4^{-1/2}f_4\,dt^2
 +\mathcal H_4^{1/2}\left(\frac{dr^2}{f_4}+r^2d\Omega_2^2\right),\\
 A_i={}&\frac{\sqrt{q_i(q_i+2m)}}{2(r+q_i)}\,dt,\qquad
 X_i=H_i^{-1}\mathcal H_4^{1/4},
\end{aligned}
\label{eq:4d-parent-fields}
\end{equation}
with
\begin{equation}
 f_4=1-\frac{2m}{r}+\frac{r^2}{\ell^2}\mathcal H_4,\qquad
 H_i=1+\frac{q_i}{r},\qquad
 \mathcal H_4=\prod_{i=1}^{4}H_i.
 \label{eq:4d-parent-functions}
\end{equation}
Here \rev{$t$ and $r$ are the time and radial coordinates, $f_4$ is the four-dimensional radial metric function, and $H_i$ and $\mathcal H_4$ are the charge-dependent functions defined above.} $d\Omega_2^2$ is the metric on the unit two-sphere, $A_i$ are the gauge potentials, $\ell$ is the AdS curvature radius, and $X_i$ are the scalar fields. In four dimensions, $P=3/(8\pi\ell^2)$; the horizon radius obeys $f_4(r_h)=0$. The EMDA solution \cite{PRL69-1006} corresponds to $q_1=q_2=q$ and $q_3=q_4=0$. Its thermodynamic quantities are
\begin{equation}
\begin{gathered}
 M=m+\frac q2,\\
 S=\pi r_h(r_h+q),\\
 T_H=\frac{1+\frac{1}{\ell^2}(r_h+q)(3r_h+q)}{4\pi(r_h+q)}.
\end{gathered}
\label{eq:emda-data}
\end{equation}
Fixing the physical charge requires fixing
\begin{equation}
 K=q(q+2m)=q[q+r_h+\frac{1}{\ell^2} r_h(r_h+q)^2].
 \label{eq:emda-constraint}
\end{equation}
\rev{In this single-charge notation, $K$ denotes the nonzero charge combination $K_i$ introduced above.} Consequently $q=q(r_h)$ is determined by Eq.~(\ref{eq:emda-constraint}), rather than held constant. Differentiating gives
\begin{equation}
 \frac{dq}{dr_h}=
 -\frac{q[1+\frac{1}{\ell^2}(r_h+q)(q+3r_h)]}
 {2q+r_h+\frac{1}{\ell^2} r_h(r_h+q)(3q+r_h)}.
 \label{eq:emda-qprime}
\end{equation}
In particular, as $r_h\to0$, $q\to\sqrt K$, and the boundary temperature $T_\partial\equiv\lim_{r_h\to0}T_H(r_h)$ is
\begin{equation}
 T_{\partial}^{\rm EMDA}=\frac{1+K/\ell^2}{4\pi\sqrt K}.
 \label{eq:emda-boundaryT}
\end{equation}

\begin{revblock}
\noindent The fixed-charge constraint in Eq.~(\ref{eq:emda-constraint}) also gives
\begin{equation}
 \left.\frac{dS}{dr_h}\right|_{K,P}^{\rm EMDA}
 =\frac{2\pi(r_h+q)^2[1+r_h(r_h+q)/\ell^2]}
 {2q+r_h+r_h(r_h+q)(3q+r_h)/\ell^2}>0.
 \label{eq:emda-entropy-check}
\end{equation}
\end{revblock}

\subsubsection{Four-dimensional Horowitz-Sen black hole}

For the Horowitz-Sen (HS) solution \cite{PRD53-808,NPB477-449}, the same four-charge parent solution is truncated to $q_1,q_2>0$ and $q_3=q_4=0$. The thermodynamic quantities are
\begin{equation}
\begin{gathered}
 M=\frac c2+\frac{q_1+q_2}{4},\qquad
 S=\pi r_h\sqrt{(r_h+q_1)(r_h+q_2)},\\
 T_H=\frac{1+\frac{1}{\ell^2}[3r_h^2+2r_h(q_1+q_2)+q_1q_2]}
 {4\pi\sqrt{(r_h+q_1)(r_h+q_2)}}.
\end{gathered}
\label{eq:hs-data}
\end{equation}
where $c=2m=r_h[1+\frac{1}{\ell^2}(r_h+q_1)(r_h+q_2)]$.
At fixed physical charges, $K_i=q_i(q_i+c)$ are constants. The charge parameters are thus determined by
\begin{equation}
 q_i(c)=\frac{\sqrt{c^2+4K_i}-c}{2},
 \qquad i=1,2,
 \label{eq:hs-constraints}
\end{equation}
together with the horizon relation defining $c$ above. Since $q_i\to\sqrt{K_i}$ as $r_h\to0$, the limiting temperature is
\begin{equation}
 T_{\partial}^{\rm HS}=
 \frac{1+\sqrt{K_1K_2}/\ell^2}
 {4\pi(K_1K_2)^{1/4}}.
 \label{eq:hs-boundaryT}
\end{equation}

\begin{revblock}
\noindent To check the entropy derivative at fixed $K_1$ and $K_2$, set $A_i=r_h+q_i$ and $b_i=q_i/(c+2q_i)$, with $0<b_i<1/2$. The constraints imply $dq_i/dr_h=-b_i\,dc/dr_h$. Introducing $D=1+ r_h(b_1A_2+b_2A_1)/\ell^2>0$ and $V=b_1A_2+b_2A_1$, one obtains
\begin{equation}
\begin{aligned}
 \left.\frac{dS}{dr_h}\right|_{K_i,P}^{\rm HS}
 &=\frac{\pi}{2\sqrt{A_1A_2}\,D}\bigl[2A_1A_2\\
 &\quad+r_h(A_1+A_2-V)+r_hA_1A_2V/\ell^2\bigr]>0.
\end{aligned}
 \label{eq:hs-entropy-check}
\end{equation}
Here $A_1+A_2-V>0$ follows from $0<b_i<1/2$.
\end{revblock}

\begin{figure*}[t]
\centering
\includegraphics[width=0.94\textwidth]{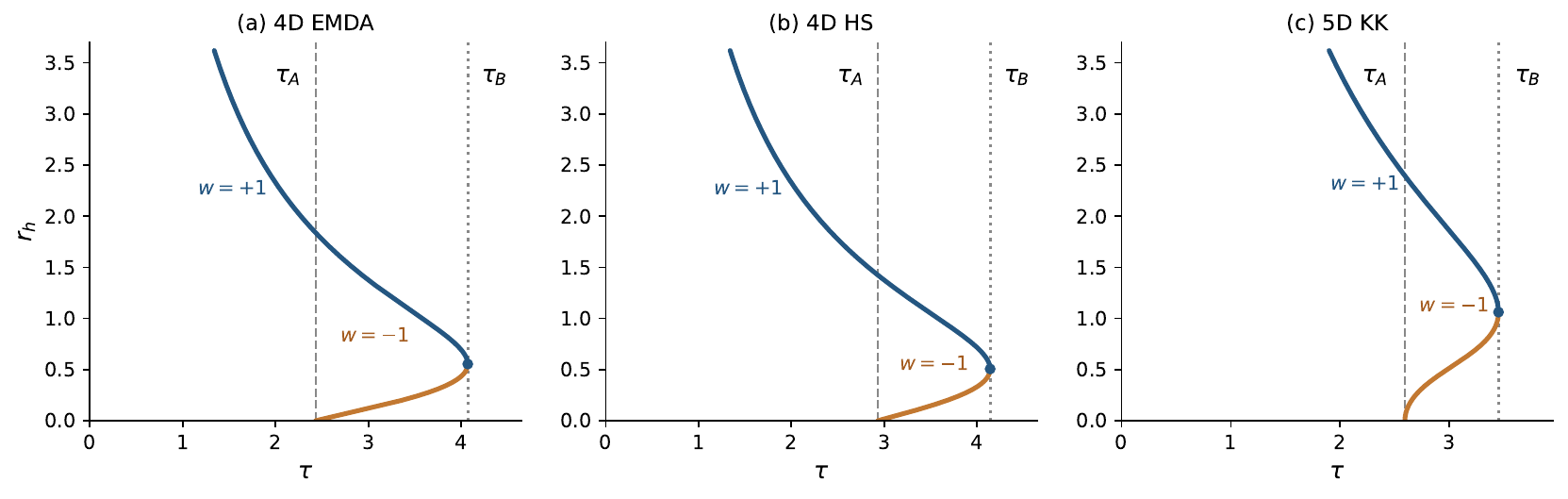}
\caption{Zero-point curves in the $(\tau,r_h)$ plane for the fixed-physical-charge families at $Pr_0^2=0.1$, where $r_0$ is a reference length: (a) four-dimensional EMDA, (b) four-dimensional HS, and (c) five-dimensional KK. The dashed vertical line at $\tau_A=1/T_\partial$ marks the boundary inverse temperature, while the dotted vertical line at $\tau_B$ passes through the maximum of the zero-point curve, corresponding to an interior pairwise merger of equilibrium roots. The upper blue branch has $w=+1$, whereas the lower orange branch has $w=-1$. The two characteristic inverse temperatures have different roles: the global topological number changes across $\tau_A$, whereas the pairwise merger at $\tau_B$ leaves it unchanged.}
\label{fig:fixed-charge-slices}
\end{figure*}

\begin{table*}[t]
\caption{Equilibrium radii $r_i$ and local winding numbers $w_i$ on two temperature slices at fixed physical charges and $Pr_0^2=0.1$ (with $r_0=1$ in the numerical units). The boundary temperatures are $T_\partial^{\rm EMDA}=0.411221$, $T_\partial^{\rm HS}=0.341204$, and $T_\partial^{\rm KK}=0.385696$. The pairs $(r_i,w_i)$ are obtained by solving the fixed-charge constraints and $T_H(r_i)=T_b$.}
\label{tab:fixedQroots}
\begin{ruledtabular}
\begin{tabular}{lcccc}
Black hole & Fixed charge parameters & $T_b/T_\partial$ & $(r_i,w_i)$ & $W$\\
\hline
EMDA & $K=0.04$ & $0.98$ & $(0.01050,-1),\ (1.79164,+1)$ & $0$\\
 & & $1.02$ & $(1.88496,+1)$ & $1$\\
HS & $K_1=0.04,\ K_2=0.09$ & $0.98$ & $(0.01468,-1),\ (1.38165,+1)$ & $0$\\
 & & $1.02$ & $(1.46647,+1)$ & $1$\\
KK & $K=0.04$ & $0.98$ & $(0.18553,-1),\ (2.32059,+1)$ & $0$\\
 & & $1.02$ & $(2.46674,+1)$ & $1$
\end{tabular}
\end{ruledtabular}
\end{table*}

\subsubsection{Five-dimensional Kaluza-Klein black hole}

The five-dimensional Kaluza-Klein (KK) example is obtained from the static three-charge AdS black hole in gauged supergravity \cite{NPB553-317}. The parent solution takes the form
\begin{equation}
\begin{aligned}
 ds_5^2={}&-\mathcal H_5^{-2/3}f_5\,dt^2
 +\mathcal H_5^{1/3}\left(\frac{dr^2}{f_5}+r^2d\Omega_3^2\right),\\
 A_i={}&\frac{\sqrt{q_i(q_i+2m)}}{r^2+q_i}\,dt,\qquad
 X_i=H_i^{-1}\mathcal H_5^{1/3},
\end{aligned}
\label{eq:5d-parent-fields}
\end{equation}
where
\begin{equation}
 f_5=1-\frac{2m}{r^2}+\frac{r^2}{\ell^2}\mathcal H_5,\qquad
 H_i=1+\frac{q_i}{r^2},\qquad
 \mathcal H_5=\prod_{i=1}^{3}H_i.
 \label{eq:5d-parent-functions}
\end{equation}
Here $d\Omega_3^2$ is the metric on the unit three-sphere, \rev{$f_5$ is the five-dimensional radial metric function, and $H_i$ and $\mathcal H_5$ are its charge-dependent functions.} $\ell$ is the AdS curvature radius, and $f_5(r_h)=0$ determines the horizon radius. In five dimensions, $P=3/(4\pi\ell^2)$. The KK solution used below is the single-charge truncation $q_1=q$, $q_2=q_3=0$. A more general four-charge extension of the five-dimensional AdS solution has also been studied in Ref.~\cite{EPJC86-187,JHEP0226252}. The thermodynamic quantities of the single-charge KK black hole are
\begin{equation}
\begin{gathered}
 M=\frac{3\pi m}{4}+\frac{\pi q}{4},\\
 S=\frac{\pi^2 x\sqrt{x+q}}{2},\qquad
 T_H=\frac{1+\frac{1}{\ell^2}(2x+q)}{2\pi\sqrt{x+q}}.
\end{gathered}
\label{eq:kk-data}
\end{equation}
where $x=r_h^2$ and $2m=x[1+\frac{1}{\ell^2}(x+q)]$.
The fixed-charge constraint is
\begin{equation}
 K=q(q+2m)=q[q+x+\frac{1}{\ell^2} x(x+q)],
 \label{eq:kk-constraint}
\end{equation}
which determines the positive branch
\begin{equation}
 q(r_h)=\frac{\sqrt{x^2+4K/(1+x/\ell^2)}-x}{2}.
 \label{eq:kk-q}
\end{equation}
At the small-radius boundary $q\to\sqrt K$, giving
\begin{equation}
 T_{\partial}^{\rm KK}=
 \frac{1+\sqrt K/\ell^2}{2\pi K^{1/4}}.
 \label{eq:kk-boundaryT}
\end{equation}

\begin{revblock}
\noindent For the KK solution, differentiating the fixed-$K$ constraint with respect to $x=r_h^2$ yields
\begin{equation}
 \left.\frac{dq}{dx}\right|_{K,P}
 =-\frac{q(1+q/\ell^2+2x/\ell^2)}
 {(1+x/\ell^2)(2q+x)}.
 \label{eq:kk-charge-check}
\end{equation}
Consequently,
\begin{equation}
 \left.\frac{dS}{dx}\right|_{K,P}^{\rm KK}
 =\frac{\pi^2\sqrt{x+q}\,[4q+3x+3x(x+q)/\ell^2]}
 {4(1+x/\ell^2)(2q+x)}>0.
 \label{eq:kk-entropy-check}
\end{equation}
Since $dx/dr_h=2r_h>0$, this also establishes $dS/dr_h>0$. \rev{Direct differentiation of the mass expressions for EMDA, HS, and KK, with the same fixed-charge constraints, verifies Eq.~(\ref{eq:fixedQ-firstlaw-check}) for all three models.}
\end{revblock}

\begin{revblock}
\subsubsection{Physical domain and the small-radius boundary}

The range $r_h>0$ corresponds to an admissible branch of finite-radius black holes in each of the three models. At fixed positive $K_i$ and $P$, the EMDA and KK charge constraints determine a unique positive $q$ for every $r_h>0$; the HS constraints similarly determine positive $q_1,q_2$ through the horizon equation. In all cases the mass parameter $m$ is positive. For the two four-dimensional truncations and the five-dimensional single-charge solution, the metric functions $f_4$ and $f_5$ (whose subscripts indicate spacetime dimension) and their radial derivatives at fixed solution parameters are
\begin{equation}
\begin{aligned}
 f_4(r)&=1-\frac{2m}{r}+\frac{(r+q_1)(r+q_2)}{\ell^2},\\
 \partial_r f_4(r)&=\frac{2m}{r^2}+\frac{2r+q_1+q_2}{\ell^2}>0,\\
 f_5(r)&=1-\frac{2m}{r^2}+\frac{r^2+q}{\ell^2},\\
 \partial_r f_5(r)&=\frac{4m}{r^3}+\frac{2r}{\ell^2}>0.
\end{aligned}
\label{eq:physical-domain}
\end{equation}
Thus each positive root of $f_4(r_h)=0$ or $f_5(r_h)=0$ is the unique positive-radius horizon of the corresponding geometry. Along the fixed-charge branches, the limiting behavior as $r_h\to0^+$ is
\begin{equation}
 m\to0,\qquad q_i\to\sqrt{K_i},\qquad
 S\to0,\qquad T_H\to T_\partial>0.
 \label{eq:zero-area-domain}
\end{equation}
More specifically, $S=O(r_h)$ for the EMDA and HS solutions, whereas $S=O(r_h^2)$ for the KK solution. The endpoint $r_h=0$ therefore lies at the zero-area boundary of the finite-radius black-hole domain. Its finite limiting Hawking temperature does not describe a regular zero-temperature extremal horizon. In this setting, $T_\partial$ marks a boundary threshold of the thermodynamic topological count; its identification does not require a conventional bulk thermodynamic critical point.
\end{revblock}

\subsubsection{Temperature-dependent topological numbers}

Let $T_b=1/\tau$ denote the temperature of a chosen slice, and $T_\partial$ the small-radius limiting Hawking temperature. For each of these fixed-charge families, $T_H$ approaches the finite value $T_\partial$ as $r_h\to0$ and diverges to $+\infty$ as $r_h\to\infty$. The entropy increases along the physical fixed-charge branches. Thus, away from degenerate equilibrium roots and the boundary slice itself, the signed intersection count of $T_H(r_h)$ with a horizontal temperature slice yields
\begin{equation}
 W(T_b)=
 \begin{cases}
 0,&0<T_b<T_\partial,\\
 1,&T_b>T_\partial.
 \end{cases}
 \label{eq:three-model-W}
\end{equation}
This result does not require the Hawking temperature to be monotonic: any additional interior intersections occur with canceling winding numbers.

As an explicit illustration, Fig.~\ref{fig:fixed-charge-slices} and Table~\ref{tab:fixedQroots} show the equilibrium radii and their winding numbers for $Pr_0^2=0.1$, where $r_0$ is an arbitrary reference length and we use $r_0=1$ as the numerical length unit. The fixed charge parameters are chosen separately for each solution. In every case the global topological number changes from zero to one across $T_\partial$, even though the physical charges and pressure are unchanged.

\begin{revblock}
The complete equilibrium-root counts clarify the meaning of $W=0$. For the specific charge choices of Table~\ref{tab:fixedQroots}, each temperature curve has one nondegenerate minimum at $T_{\min}<T_\partial$, with $T_{\min}=0.245648$, $0.241487$, and $0.289928$ for EMDA, HS, and KK, respectively. Accordingly, for ordinary temperature slices away from the two special values,
\begin{equation}
 N_{\rm eq}(T_b)=
 \begin{cases}
 0,&0<T_b<T_{\min},\\
 2,&T_{\min}<T_b<T_\partial,\\
 1,&T_b>T_\partial.
 \end{cases}
 \label{eq:root-count-regimes}
\end{equation}
\rev{Here $N_{\rm eq}$ is the number of positive-radius equilibrium black holes on the temperature slice $T_b$.}
The two roots in the middle interval have winding numbers $-1$ and $+1$ and therefore give $W=0$, although both are genuine finite-radius equilibrium black holes. The transition at $T_\partial$ thus connects two temperature regimes that both contain equilibrium black holes. The corresponding inverse temperatures satisfy $\tau_A=1/T_\partial<\tau_B=1/T_{\min}$, as shown in Fig.~\ref{fig:fixed-charge-slices}.
\end{revblock}

These examples establish temperature-dependent changes in the global thermodynamic topological number within the fixed-physical-charge description. In the next section, we examine why these changes are associated with the small-radius boundary, rather than with the creation or annihilation of pairs of interior equilibrium states.

\section{Physical Mechanisms of Temperature-Dependent Topological Phase Transitions}\label{III}

The three examples in Sec.~\ref{II} exhibit a change in the global topological number while the physical charges and pressure remain fixed. We now distinguish the two types of events visible in their zero-point curves: the passage of an equilibrium root through the boundary at $r_h=0$ and the merger of two roots in its interior. Although both alter the number of equilibrium states on a given temperature slice, only the former changes their total winding number.

\subsection{Fixed-charge equilibrium curves and local winding numbers}

Let $r_h$ denote a coordinate on a physical fixed-charge branch, with $r_h>0$, and let $T_b=1/\tau>0$. Along this branch the first law gives
\begin{equation}
 \left.\frac{d\mathcal F}{dr_h}\right|_{Q_i,P}
 =S'(r_h)[T_H(r_h)-T_b],
 \label{eq:mechanism-Fprime}
\end{equation}
where a prime denotes differentiation along the fixed-charge family. Thus an equilibrium root $r_i$ satisfies $T_H(r_i)=T_b$. At a simple root,
\begin{equation}
 \left.\frac{d\phi^{r_h}}{dr_h}\right|_{r_i}
 =S'(r_i)T_H'(r_i).
 \label{eq:mechanism-jacobian}
\end{equation}
Since $\partial_\Theta(-\cot\Theta\csc\Theta)|_{\Theta=\pi/2}=1$, and $S'(r_i)>0$ on the branches under consideration, the local winding number is
\begin{equation}
 w_i=\operatorname{sgn}T_H'(r_i)
 =-\operatorname{sgn}\tau'(r_i),
 \qquad \tau(r_h)=\frac{1}{T_H(r_h)}.
 \label{eq:mechanism-winding}
\end{equation}
This relation connects the winding numbers directly to the orientations of the branches in Fig.~\ref{fig:fixed-charge-slices}. In particular, the upper and lower branches on the displayed examples carry $w=+1$ and $w=-1$, respectively.

\subsection{Boundary crossing and the change of the global topological number}

The decisive feature shared by the three fixed-charge examples is the finite small-radius limiting temperature,
\begin{equation}
 \lim_{r_h\to0}T_H(r_h)=T_\partial>0,
 \qquad
 \lim_{r_h\to\infty}T_H(r_h)=+\infty.
 \label{eq:mechanism-endpoints}
\end{equation}
The explicit values of $T_\partial$ are given in Eqs.~(\ref{eq:emda-boundaryT}), (\ref{eq:hs-boundaryT}), and (\ref{eq:kk-boundaryT}). At a regular temperature $T_b$ different from these endpoint values, the signed number of crossings of $T_H(r_h)-T_b$ can be evaluated from its endpoint signs:
\begin{equation}
 \begin{split}
 W(T_b)
 &=\sum_{T_H(r_i)=T_b}\operatorname{sgn}T_H'(r_i)\\
 &=\frac12\left[\operatorname{sgn}\bigl(T_H(\infty)-T_b\bigr)
 -\operatorname{sgn}\bigl(T_\partial-T_b\bigr)\right].
 \end{split}
 \label{eq:mechanism-crossing}
\end{equation}
Here $T_H(\infty)=+\infty$ is understood as a limiting statement. Equation~(\ref{eq:mechanism-crossing}) follows because successive simple crossings of a continuous one-dimensional function alternate in orientation; their signed sum depends only on the signs at the two ends. It therefore gives $W=0$ for $0<T_b<T_\partial$ and $W=1$ for $T_b>T_\partial$, independently of any additional interior turning points.

\begin{revblock}
More generally, Eq.~(\ref{eq:mechanism-crossing}) does not require the equilibrium curve to have only one interior turning point, as in Fig.~\ref{fig:fixed-charge-slices}. The same continuous fixed-charge temperature curve may have arbitrarily many isolated folds. A temperature slice can then intersect the curve at additional equilibrium radii, but these extra roots appear or disappear in pairs with opposite winding numbers, contributing $(+1)+(-1)=0$ to $W$. The number and distribution of equilibrium black holes can therefore vary considerably between temperature slices, while the total topological number remains controlled by the two ends of the same equilibrium curve. This independence from the detailed interior structure is the key physical content of Eq.~(\ref{eq:mechanism-crossing}).
\end{revblock}

In the $(\tau,r_h)$ plane, the corresponding boundary value is
\begin{equation}
 \tau_A=\frac{1}{T_\partial}.
 \label{eq:mechanism-tauA}
\end{equation}
As $\tau$ crosses $\tau_A$, an equilibrium root reaches or emerges from the open boundary $r_h=0$. The signed intersection count can consequently change by one. This is a boundary event, not a merger of two equilibrium roots at a positive horizon radius. \rev{As established by Eq.~(\ref{eq:zero-area-domain}), this crossing occurs at the zero-area endpoint of a family of genuine black holes, while the remaining equilibrium roots retain strictly positive horizon radii.}

\subsection{Interior pairwise merger and conservation of the winding number}

A different event occurs at an interior turning point $r_B>0$ of the zero-point curve, where
\begin{equation}
 \tau'(r_B)=0,
 \qquad \tau_B=\tau(r_B),
 \label{eq:mechanism-tauB}
\end{equation}
provided the turning point is nondegenerate, $\tau''(r_B)\ne0$. Locally the equilibrium condition has the expansion
\begin{equation}
 \tau-\tau_B=\frac12\tau''(r_B)(r_h-r_B)^2
 +O\!\left((r_h-r_B)^3\right).
 \label{eq:mechanism-fold}
\end{equation}
On one side of $\tau_B$ this equation has two nearby simple roots; on the other side it has none. The derivatives $\tau'(r_i)$ at the two roots have opposite signs. Equation~(\ref{eq:mechanism-winding}) therefore gives
\begin{equation}
 w_++w_-= (+1)+(-1)=0.
 \label{eq:mechanism-cancellation}
\end{equation}
\rev{Here $w_+$ and $w_-$ denote the winding numbers of the two roots involved in the interior pairwise event.}
The pairwise creation or annihilation changes the number of equilibrium states but not the global topological number. Exactly at the merger the simple-root winding formula is inapplicable, but the signed count on either side is unchanged.

The two characteristic values marked in Fig.~\ref{fig:fixed-charge-slices} consequently have distinct meanings. The boundary value $\tau_A$ separates temperature slices with different $W$, whereas the interior turning-point value $\tau_B$ changes the number of roots without changing $W$. Additional interior pairwise events likewise leave the signed count unchanged.

\subsection{Implications for the three charged AdS black holes}

The model dependence enters through the fixed-charge constraints and the resulting Hawking temperature functions. For the EMDA black hole, $q(r_h)$ is determined by Eq.~(\ref{eq:emda-constraint}); for the HS black hole, both $q_i(r_h)$ follow from Eq.~(\ref{eq:hs-constraints}) and the horizon relation; and for the five-dimensional KK black hole, $q(r_h)$ follows from Eq.~(\ref{eq:kk-constraint}). These different constraints produce different values of $T_\partial$ and different locations of interior turning points, but the mechanism governing their total winding numbers is the same.

In particular, the fixed-charge calculations establish that the entropy derivatives are positive on the positive-charge branches considered, and the limiting behaviors in Eq.~(\ref{eq:mechanism-endpoints}) hold. These are the ingredients required by Eq.~(\ref{eq:mechanism-crossing}). The explicit root counts in Table~\ref{tab:fixedQroots} illustrate the mechanism: below the boundary temperature the two roots carry opposite winding numbers, while above it the remaining root carries winding number $+1$. The disappearance of the negative-winding root through the small-radius boundary accounts for the change in $W$; an interior pairwise merger, by contrast, cannot account for such a change.

This distinction provides the physical connection sought here. The different topological numbers found on different temperature slices are determined by the same fixed-charge equilibrium curve and its small-radius boundary. We use this result in Sec.~\ref{IV} to discuss the thermodynamic projection conjecture.

\section{Thermodynamic Projection Conjecture}\label{IV}

The results presented in Secs.~\ref{II} and \ref{III} demonstrate that different thermodynamic topological outcomes can emerge from the same underlying black hole thermodynamic structure. These different outcomes are not independent of one another, but are connected through the equilibrium structure and the physical mechanisms governing the evolution of equilibrium states. From this perspective, the topological properties observed under different thermodynamic conditions can be understood as different manifestations of a common underlying thermodynamic structure.

This observation motivates us to propose the \emph{thermodynamic projection conjecture}:

\begin{quote}
\emph{Different thermodynamic topological manifestations of a given black hole system can be understood as projections of its complete thermodynamic structure under different thermodynamic conditions. Their differences and underlying connections are determined by the structure of the equilibrium states and the physical mechanisms governing their evolution.}
\end{quote}

In the three black hole models considered here, the change in $W$ occurs when an equilibrium state crosses the boundary at $r_h=0$, whereas the pairwise creation or annihilation of equilibrium states at finite horizon radii leaves $W$ unchanged. At a given temperature, \rev{on the fixed-charge branches with $S' > 0$ considered here,} the global topological number $W$ represents the difference between the numbers of locally thermodynamically stable and unstable equilibrium black holes. As demonstrated by Eq.~(\ref{eq:mechanism-crossing}) derived in Sec.~\ref{III}, its value and variation can be directly connected to the boundary behavior of the underlying equilibrium structure.

\begin{revblock}
The connection between these physical mechanisms and the thermodynamic projection conjecture can be expressed more explicitly. At fixed physical charges $Q_i$ and pressure $P$, the Hawking temperature $T_H(r_h)$ specifies the complete equilibrium curve considered here. Changing the external bath temperature $T_b$ selects different equilibrium radii through $T_H(r_i)=T_b$, but leaves the underlying curve unchanged. \rev{Here $W$ refers to the selected temperature slice, with $Q_i$ and $P$ held fixed.} Combining the local winding-number relation in Eq.~(\ref{eq:mechanism-winding}) with the boundary result in Eq.~(\ref{eq:mechanism-crossing}), we can write
\begin{equation}
\begin{split}
W
 &=\sum_{T_H(r_i)=T_b}\operatorname{sgn}T_H'(r_i)\\
 &=\frac12\left[\operatorname{sgn}\bigl(T_H(\infty)-T_b\bigr)
  -\operatorname{sgn}\bigl(T_\partial-T_b\bigr)\right].
\end{split}
\label{eq:projection-physical-link}
\end{equation}
The first line uses the winding numbers of the individual equilibrium black holes; the second line restates Eq.~(\ref{eq:mechanism-crossing}) in terms of the boundary temperatures of their common equilibrium curve. The difference between these two descriptions is precisely what the temperature-projection picture emphasizes: the winding numbers are obtained from individual equilibrium states on a temperature slice, while their sum is fixed by the global boundary behavior.

In particular, for any two regular bath temperatures $T_1,T_2>0$ at the same $Q_i$ and $P$, subtraction of Eq.~(\ref{eq:projection-physical-link}) gives
\begin{equation}
\begin{split}
W(T_2)-W(T_1)=\frac12\bigl[\operatorname{sgn}(T_\partial-T_1)
 -\operatorname{sgn}(T_\partial-T_2)\bigr].
\end{split}
\label{eq:projection-between-slices}
\end{equation}
Thus two temperature slices on the same side of $T_\partial$ have the same global topological number, even if the equilibrium curve folds repeatedly between them and the numbers of equilibrium roots differ. When the slices lie on opposite sides of $T_\partial$, their global topological numbers differ by one, reproducing Eq.~(\ref{eq:three-model-W}). As demonstrated by Eq.~(\ref{eq:mechanism-cancellation}), any additional interior fold creates or annihilates roots with cancelling winding contributions. These equations therefore connect the different temperature-dependent topological manifestations directly through the boundary and interior mechanisms of the same black hole equilibrium structure.
\end{revblock}

More importantly, this perspective provides a systematic approach to investigating the mechanisms underlying thermodynamic topological phenomena by establishing their connections to the thermodynamic relations and analytical properties of the complete equilibrium structure. The temperature-dependent topological transitions examined in this work provide a representative example of this approach, as both the changes in the global topological number and the conservation of winding numbers during interior pairwise processes follow from the same fixed-charge equilibrium structure. \rev{Exploring whether analogous connections organize other thermodynamic conditions and topological classes is the broader content of the thermodynamic projection conjecture.}

\section{Conclusions and Outlook}\label{V}

The successive development of thermodynamic topological classifications has considerably enriched the description of black hole equilibrium states. Nevertheless, as the distinctions between classes and subclasses become increasingly detailed, the additional physical insight gained from further classification becomes less evident. Identifying more topological patterns alone does not establish why they arise or how they are related within the thermodynamics of a black hole system. This motivates us to move beyond the classification of separate outcomes and seek a more holistic understanding of their common physical origin. In this work, we have therefore proposed the \emph{thermodynamic projection conjecture}, which views different thermodynamic topological manifestations as arising from the complete thermodynamic structure of a given black hole system under different thermodynamic conditions. The central idea is that their distinctions and connections should be understood through the equilibrium structure and the mechanisms governing its evolution.

To investigate this idea, we have used temperature-dependent thermodynamic topological phase transitions as a representative example, considering the four-dimensional Einstein-Maxwell-dilaton-axion and Horowitz-Sen black holes and the five-dimensional Kaluza-Klein black hole. With the physical charges and pressure held fixed, we have established that the different global topological numbers observed at different temperatures are determined by the same underlying equilibrium structure. In particular, we have derived an analytical relation between the global topological number and the boundary behavior of the Hawking temperature. This relation identifies the crossing of an equilibrium state through $r_h=0$ as the mechanism responsible for changes in the global topological number. In contrast, the creation or annihilation of equilibrium states in pairs at finite horizon radii does not alter this number, because their winding contributions cancel. The two mechanisms therefore account for different aspects of the evolution of equilibrium states within a single thermodynamic structure.

These results provide analytical support for the thermodynamic projection conjecture by demonstrating how different topological outcomes can be traced to a common equilibrium structure and its underlying physical mechanisms. A natural direction for further investigation is to examine the four previously established thermodynamic topological classes and their associated subclasses within this framework, with particular emphasis on the physical mechanisms underlying their differences and connections. Furthermore, the thermodynamic projection perspective we have provided offers a new guiding principle for investigating other thermodynamic systems, in which seemingly distinct phenomena can be understood as interconnected manifestations of a common underlying thermodynamic structure.

\begin{acknowledgments}
This work is supported by the National Natural Science Foundation of
China (NSFC) under Grants No. 12675077, No. 12205243 and No. 12375053, by the China
Scholarship Council (CSC), and by the Sichuan Science and Technology Program
under Grant No. 2026NSFSC0021.
\end{acknowledgments}

\end{document}